\documentclass[12pt,preprintnumbers]{article}
\usepackage{amsmath, amssymb, wasysym, slashed, multirow,hyperref}
\usepackage{pdfpages}
\usepackage{cite}
\usepackage{dsfont}
\usepackage{times}
\usepackage{xcolor}
\usepackage{graphicx,color}  %
\usepackage{bm}  %
\usepackage{enumitem}
\newenvironment{sciabstract}{%
\begin{quote} }
{\end{quote}}

\newcounter{lastnote}

\usepackage{fancyhdr}
\fancypagestyle{plain}{%
	\fancyhead[R]{}

}

\title{Resurgence and self-completion in renormalized gauge theories}

\author
{Alessio Maiezza,$^{1\ast}$  Juan Carlos Vasquez$^{2\dagger}$\\
\\
\normalsize{$^{1}$Dipartimento di Scienze Fisiche e Chimiche, Universit\`a} \\ \\
\normalsize{degli Studi dell'Aquila, via Vetoio, I-67100, L'Aquila, Italy,}\\ \\
\normalsize{$^{2}$Department of Physics $\&$ Astronomy, Amherst College, Amherst, MA 01002, USA} \\
\\
\small{ E-mail: alessiomaiezza@gmail.com$^{\ast}$, jvasquezcarmona@amherst.edu$^{\dagger}$}
}

\date{}

\begin{document}


\baselineskip16pt 


\maketitle


\begin{sciabstract}
Under certain assumptions and independent of the instantons, we show that the logarithm expansion of dimensional regularization in quantum field theory needs a nonperturbative completion to have a renormalization-group flow valid at all energies. Then, we show that such nonperturbative completion has the analytic properties of the renormalons, which we find with only a marginal reference to diagrammatic calculations. We demonstrate that renormalon corrections necessarily lead to analyzable functions, namely, resurgent transseries. A detailed analysis of the resurgent properties of the renormalons is provided. The self-consistency of the theory requires these nonperturbative contributions to render the running coupling well-defined at any energy, thus with no Landau pole. We illustrate the point within the case of QED. This way, we explicitly realize the correspondence between the nonperturbative Landau pole scale and the renormalons. What is seen as a Landau pole in perturbation theory is cured by the nonperturbative, resurgent contributions.
\end{sciabstract}

\section{Introduction}\label{sec:introduction}

Power series, coming from perturbation theory in quantum field theory (QFT), have been known to be, at best, asymptotic~\cite{Dyson:1952tj} and non-Borel-Laplace resummable. One reason is that the series has a $n!$ behavior for large $n$, being this the order of perturbation theory. The sources of the $n!$ behavior are the instantons and the renormalons. The former objects are understood in terms of the growing number of Feynman diagrams: one has $\mathcal{O}(n!)$ diagrams at order $n$, for $n$ sufficiently large. The $n!$ behavior can lead to singularities on the integration path of the Laplace integral $f(x) =\int_0^{\infty} B(z) \, e^{-zx}  \, dz\,$, thus preventing summability. However, since the instantons singularities can be traced back to the minimization of the classical action, they have a semiclassical limit which, in principle, can circumvent the problem~\cite{Lipatov:1976ny} -- see Ref.~\cite{zinn2011barrier} for a review, and Refs.~\cite{Broadhurst:1999ys,Borinsky:2021hnd} for
applications of Hopf algebraic approach to Schwinger–Dyson equation. Thus, the instantons do not pose a foundational problem for perturbation theory~\cite{tHooft:1977xjm}.

A different story holds for the renormalons~\cite{Gross:1974jv,Lautrup:1977hs,tHooft:1977xjm,Parisi:1978iq,Neubert:1994vb}. They are originated only from a few diagrams, $\mathcal{O}(1)$, which give the $n!$ large order behavior and, possibly, ambiguities into the Laplace integral not related to any semiclassical limit. Therefore, the renormalons represent a severe problem for the consistency of perturbative QFT. On the other hand, they may give an estimate of large order behavior~\cite{Beneke:1998ui}, motivating observable evaluations in Quantum Chromodynamics~\cite{Cvetic:2019jmu,Caprini:2020lff,Ayala:2021mwc,Ayala:2022cxo,Caprini:2023tfa}. Although the current literature has several different applications (e.g. Refs.~\cite{Correa:2019xvw,Loewe:2021ekj,Loewe:2022aaw} for magnetic renormalons and finite temperature applications) and estimates for renormalons(e.g. Ref.~\cite{deCalan:1981szv}), based on the so-called ('t Hooft) skeleton diagrams, some authors question whether these particular diagrams might cancel out -- see the discussions in Ref.~\cite{Suslov:1999cg}. In other words, there is the feeling in the literature that the skeleton-diagrams-based arguments are a poor foundation for the existence of the renormalons, even though there are compelling pieces of evidence of renormalons from numerical simulations in Quantum Chromodynamics (QCD)~\cite{Bauer:2011ws}. By ``existence" of the renormalons, we mean that the kind of Feynman diagrams typically associated with renormalons do not cancel and contribute factorially.

One central scope of the present article is to provide a generic argument for the existence of renormalons in QFT, with no direct reference to skeleton diagrams calculations, but only relying on the one-loop beta function. The fundamental \emph{agreement} assumed in this work is that the Callan-Symanzik equation or Renormalization Group Equation (RGE) is \textit{per se} nonperturbative. Therefore, it works beyond the logic of order by order in perturbation theory. Let us briefly anticipate our main findings. First, we show the limitations of perturbation theory, regardless of the instantons. Then, demanding that the renormalization group flow is well-defined at any energy, we show that the Green functions must be augmented with a nonanalytic function in the coupling constant.  We then show that this function has the analytic properties of renormalons and serves as a self-completion. The necessary nonperturbative self-completion, in turn, implies that the (perturbative) Landau pole must be avoided, and we illustrate how it can be achieved for $U(1)$ gauge model -- that we call Quantum Electrodynamics (QED).

The mathematical framework we shall rely on is resurgence~\cite{Ecalle1993,EcalleRes:book,Costin1995,costin1998} -- see Ref.~\cite{sauzin2007resurgent,Dorigoni:2014hea,Aniceto:2018bis} for reviews. The resurgence has recently been imported into QFT~\cite{Dunne:2013ada,Aniceto:2018uik,Maiezza:2019dht,Bersini:2019axn,Clavier:2019sph,Borinsky:2020vae,Fujimori:2021oqg,Marino:2023epd,Laenen:2023hzu,Caprini:2023kpw}. In particular, we shall elaborate on the result of Ref.~\cite{Bersini:2019axn}, in which the authors provide a nonlinear ordinary differential equation (ODE), extracted from RGE, whose solution has properties compatible with the ones of renormalons. Here, we shall show that a nonperturbative contribution to the running coupling necessarily emerges from the requirement of self-completion.

In addition, we shall sharpen the derivation of the ODE describing the renormalons with several technicalities. Among others, we shall pay particular attention to the analysis of the resurgent properties of the ODE, connecting the formalism of Ref.~\cite{CostinBook} (e.g., applied in the context of QFT in Refs.~\cite{Maiezza:2019dht}) with the one widely known in the literature, based on the Alien Calculus (e.g., Refs.~\cite{Bellon:2016mje,Borinsky:2022knn,Laenen:2023hzu}). In other words, relying on the Ecalle bridge equation, we provide an alternative derivation of the resummation isomorphism proposed in Ref.~\cite{Costin1995}.

The article is organized as follows. In Sec.~\ref{sec:RGE}, we explicitly show that the two-point correlator needs a nonperturbative contribution to make the RGE flow valid at all energies. In Sec.~\ref{sec:ODE}, we prove that such a nonperturbative completion has the analytic properties of the renormalons. Thus, it has to be identified with the Borel-Ecalle resummation of them. In Sec.~\ref{sec:resurgence}, we provide a detailed analysis of the resurgent properties of the renormalons. In Sec.~\ref{sec:QED}, we explicitly check the self-consistency of QED with a well-defined renormalization-group (RG) flow at all energies. Finally, in Sec.~\ref{sec:summary}, we conclude and give further technical details in App.~\ref{app:resum} and~\ref{app:generic}.

\section{Renormalized two-point function}\label{sec:RGE}

The scope of this preliminary section is to rephrase the perturbative Landau pole problem in a notation similar to the one adopted in Ref.~\cite{Klaczynski:2013fca}. As it is well known, the presence of the Landau pole -- seen in perturbation theory -- demands completion of the perturbative treatment.

Consider the renormalized, one-particle-irreducible two-point function valid for any massless gauge field
\begin{equation}\label{Green2}
\Pi_{\mu\nu}(q) = \left(g_{\mu\nu}-\frac{q_{\mu}q_{\nu}}{q^2}\right) \Pi(q^2) \,.
\end{equation}
Define $\mu_0^2=-q^2$ and $L=\log(\mu_0^2/\mu^2)$.  Within dimensional regularization, Perturbative renormalization expresses $\Pi$ in powers of $L$, whose coefficients are powers of $\alpha$. The latter is the all-loops correction to the tree-level propagator (or two-point function). Therefore, a generic expression for $\Pi$, or a scale expansion, can be written~\cite{KreimerYeats2006,Yeats2008,KreimerYeats2008,Kreimer2008,vanBaalen:2008tc,vanBaalen:2009hu,Klaczynski:2013fca}
\begin{equation}\label{PT_ren}
\Pi(L)= 1- \sum_{k=1}^{\infty} \gamma_k(\alpha) L^k \,,
\end{equation}
where the $\gamma_k$ have contributions of lowest order $\alpha^k$.
Notice that the finite part, i.e., the piece proportional to $L^0$, can be normalized to one order by order in perturbation theory. The Eq.~\eqref{Green2} obeys to the RGE~\cite{PhysRevD.2.1541,Symanzik:1973pp} which, in the massless limit, can be rewritten for the expression~\eqref{PT_ren} as
\begin{equation}\label{CS}
\left[-2 \partial_L +\beta(\alpha) \partial_\alpha - 2 \gamma(\alpha) \right]  \, \Pi(L) = 0 \,.
\end{equation}
Which is equivalent to an infinite system of ordinary differential equations:
\begin{align}\label{ODE_system}
& \gamma(\alpha)= \gamma_1(\alpha)  \\
& 2  \gamma(\alpha) \gamma_k(\alpha) + 2(k+1) \gamma_{k+1}(\alpha) = \beta(\alpha) \gamma_k'(\alpha) \nonumber  \,,
\end{align}
where the prime denotes derivation for $\alpha$ and $k=2,3,... , \infty$.

As anticipated above, the appearance of the Landau pole in perturbation theory is equivalent to the following  \\

\emph{Proposition 1:} Identifying the $\gamma_k$ with their truncated perturbative expressions, the power series in the variable $L$ in Eq.~\eqref{PT_ren} has a finite radius of absolute convergence $r$, which, at the leading order in $\alpha$, is
\begin{equation}\label{conv}
|L| = |\log(\mu_0^2/\mu^2)| < r= \lim_{k\rightarrow\infty} \left| \frac{\gamma_k}{\gamma_{k+1}}\right| \simeq \left|\frac{2}{\alpha \beta_1}\right|\,,
\end{equation}
where $\beta_1$ is the one-loop coefficient of the beta-function, namely $\beta(\alpha)= \beta_1 \alpha^2 + \mathcal{O}(\alpha^3)$.  For clarity, let us stress that the radius of convergence in the proposition above refers to the variable $L$. At the same time, the expansion in the coupling $\alpha$ has, as usual, zero radius of convergence.

The convergence radius in Eq.~\eqref{conv} has to be identified with the (one-loop) Landau pole. \\

The proof of the above proposition is rather straightforward. We recursively solves the system in Eq.~\eqref{ODE_system} within perturbation theory, finding at leading order in the coupling (see Appendix~\ref{app:ODEsdetail})
\begin{equation}\label{ratio}
\frac{\gamma_k(\alpha)}{\gamma_{k+1}(\alpha)} \simeq \frac{1}{\alpha \left( \frac{k \beta_1}{2(k+1)}- \frac{g_1}{k+1} \right)}\,,
\end{equation}
where $g_1$ is the one-loop coefficient of $\gamma(\alpha)$: $\gamma(\alpha)= g_1 \alpha + \mathcal{O}(\alpha)^2$. Doing the limit for $k \rightarrow \infty$, Eq.~\eqref{ratio} gives
\begin{equation}\label{r}
r \simeq \left| \frac{2}{\beta_1 \alpha} \right|\,,
\end{equation}
where the approximation symbol means to recall the leading expansion in perturbation theory.
Thus Eq.~\eqref{PT_ren} converges absolutely if
\begin{equation}\label{rbis}
|L|  < \frac{2 s}{\beta_1 \alpha} \,,
\end{equation}
where $s$  is the sign of $\beta_1$. More precisely, $s=1$ for models which are not asymptotically free (i.e. $\beta_1>0$), and $s=-1$ for asymptotically free models (i.e. $\beta_1<0$).\\

What about including higher order corrections to $\gamma_1$ and $\beta$ to Eq.~\eqref{rbis}? Given that $\alpha$ is formally small in a given energy regime,
higher loop corrections would be sub-leading, leaving unchanged the above conclusions. In addition, one might worry about the non-convergence of the series in $\alpha$ due to instantons. Treatment of these together with the renormalons is an open problem. However, since renormalons give the leading singularities in the Borel plane, and as far as one is in a regime with a sufficiently small coupling, one may expect that the instantons can be dealt with optimal truncation at order $n^*\sim 1/\alpha$.

\subsection{Renormalization Group Equation beyond perturbation theory}\label{sub:complete}

The starting point of the present analysis is that RGE holds beyond perturbation theory and, to have a well-defined RG flow of the Green functions for a fundamental system, it has to be valid at all energies. Its validity is regardless of the loop expansion. Callan-Symanzik equation is thought to hold for nonrenormalizable models -- see, for example, Ref.~\cite{Parisi:1977uz}. In other words, RGE can be taken as a \emph{principle}, complementary to the equation of motion, to deal with QFT~\footnote{In the same spirit of the Schwinger-Dyson equation approach introduced in Ref.~\cite{Schwinger1951}}. Something along this line is also implemented in the asymptotically safe scenario for quantum gravity~\cite{Weinberg:1980gg}. The concluding section shall provide further discussions on our starting point in light of known literature.

One consequence of the above \emph{principle} is as follows: beyond perturbation theory, Eq.~\eqref{CS} must hold at any energy. From Eq.~\eqref{rbis} it follows that  Eq.~\eqref{CS} is ill-defined for $L= \frac{2}{\beta_1\alpha}$. The expression~\eqref{PT_ren} can be completed with an additional function $R(\alpha)$,
explicitly defined to let the two-point correlator exist for any $L$ (or at any energy)~\footnote{Since the final result for $\Pi(q)$ after the resummation explained in the next sections, preserves the form of Eq.~\eqref{Green2}, it automatically preserves the Ward identity $q_{\mu}\Pi^{\mu\nu}(q^2)=0$ and hence gauge invariance. }:
\begin{equation}\label{complete}
\Pi(L)= 1+R(\alpha)- \sum_{k=1}^{\infty} \gamma_k (\alpha) L^k \,.
\end{equation}
In this way, $R$ changes the first equation in the system shown in ~\eqref{ODE_system}, then it propagates in the recursion, modifying Eqs.~\eqref{ratio} and~\eqref{r}.  In addition, since by definition $R$ makes  $\Pi(L)$ well-defined at all energies, which also means at  $\mu=\Lambda$ (for $\beta_1>0$),  it follows from Eq.~\eqref{rbis} that
\begin{equation}\label{Landau}
\Lambda^2= \mu_0^2 e^{\frac{2 }{\beta_1 \alpha}}\,.
\end{equation}
 Hence,  the function  $R(\alpha)$ introduces in $\Pi_{\mu\nu}$ a non-analytic dependence on the coupling constant at $\alpha =0$ -- or in QFT jargon the $R(\alpha)$ function must be nonperturbative. The latter is supported by Ref.~\cite{Balduf2023}, which shows that, in the minimal subtraction scheme, the quantity $\gamma_0(\alpha)$ is a divergent power series, thus leading to nonperturbative contributions. Furthermore, Eq.~\eqref{Landau} also suggests the transseries structure of $R$ that we shall study in the next section.

Before moving on, a few comments are in order.
We shall identify $R(\alpha)$ with the Borel-Ecalle resummation of the renormalons~\cite{Bersini:2019axn}. From here on, we shall call renormalons the singularities on the semi-positive axis in the Borel plane related to the one-loop beta function. For example, one has the renormalons in QED at
\begin{equation}\label{QEDrenexample}
z= 2 n/\beta_1 \,,
\end{equation}
being $n$ a positive integer. Notice that our definition is different from the one often read in literature: usually, renormalons are all the factorially divergent contributions related to the one-loop coefficient of the beta function, regardless of the location of the singularities in the Borel plane. While the function $R(\alpha)$ was already introduced in Ref.~\cite{Bersini:2019axn} as a reasonable hypothesis, we argue here that it must be considered for consistency. In what follows, we shall improve several aspects of the results obtained in Ref.~\cite{Bersini:2019axn}.

\section{A non-linear equation from renormalization group}\label{sec:ODE}

Starting from Eq.~\eqref{complete}, the authors in  Ref.~\cite{Bersini:2019axn} showed that $R(\alpha)$ obeys a specific nonlinear differential equation. However, in the original derivation presented in Ref.~\cite{Bersini:2019axn}, the authors use the asymptotic expression of the beta function,  hence making the completeness of their result less transparent. In this section, we present a new rigorous derivation of the result given in Ref.~\cite{Bersini:2019axn} and show their conclusions remain unchanged when considering the full, nonperturbative expression of the beta function.  To this end, let us first make a change of variable,
\begin{equation}\label{change_var}
x:= \frac{2\, s}{\beta_1 \, \alpha} \,,
\end{equation}
such that any nonanalytic dependence around the origin is moved to infinity and, we recall, $s$ is the sign of $\beta_1$ such that $x>0$. Notice that the change of variable is such that $1/x<1$ as long as $\alpha<2/\beta_1$.
Following the mathematical literature,
we shall label the set of all operations in the variable $x$ as the multiplicative model, and the set of the operations in the variable $z$, related to $x$ via Borel transforms, as the convolutive model.  For the discussion to follow, we assume the following:

\begin{itemize}
    \item
 $x$ is formally large, and one can expand for small $1/x$. The latter is equivalent to saying that perturbation theory in $\alpha$ makes sense for sufficiently small $\alpha$ in the proper energy regime where perturbation theory can be trusted.

 \item
 $R(x)$ is formally small, and one can also expand for small $R$. Thus, any subsequent rearrangement of the  RGE shown in \eqref{CS} is done for $1/x$ and $R$ small. The smallness of $R$ is equivalent to saying that the nonperturbative corrections are small in the specific energy regime, i.e., sufficiently low energy in a QED-like model or at sufficiently high energy in a pure Yang-Mills theory -- or any asymptotically free model.
 \end{itemize}

Under these assumptions, one has the following\\

\emph{Proposition 2.}
The function $R(x)$ in the convolutive model, i.e., its Borel transform  given by
\begin{equation}
B[R(x)]= \mathcal{R}(z),
\end{equation}
has an infinite number of singularities in $P$
\begin{equation}\label{singularities}
P=\{z\in \mathbb{R}, z= n    \} \,,
\end{equation}
where $n$ is a positive integer. $P$ defines a Stokes line lying on the semi-positive axis and has to be identified with the set of renormalons singularities, which lie precisely at $z=n$.
The reason is that in the multiplicative model, $R$ obeys the nonlinear ODE
\begin{align}
R'(x) &= F(1/x,R(x)) =  \nonumber \\
 &= -R(x) + k_1 \frac{R(x)}{x} +f(\mathcal{O}(R^2(x)), \mathcal{O}(R \, x^{-2})) + \text{higher order analytic terms}    \label{ODE}  \,,
\end{align}
that is in a one-to-one correspondence with the existence of renormalons. The constant $k_1$  determines the type of poles in the Borel transform of the function $R(x)$~\cite{Costin1995,costin1998,CostinBook}. We do not explicitly write the nonlinear terms of $R$ inside $f$, but notice that the presence of nonlinearity is the essential property leading to the existence of the infinite number of singularities shown in Eq.~\eqref{singularities}. \\

The proof of the above proposition, based on obtaining Eq.~\eqref{ODE}, is as follows: the Eq.~\eqref{complete} implies that the first equation of the system~\eqref{ODE_system} is modified as
\begin{equation}\label{maineq}
2 (R(x)+1) \gamma(x)=2 \gamma_1(x)-\frac{s \beta_1}{2} x^2 \beta(x) R'(x) \,.
\end{equation}
It is manifest that as $R\rightarrow 0$, one recovers $\gamma=\gamma_1$, therefore it must be that
\begin{equation}\label{gamma}
\gamma(x)= \gamma_1(x) + g\left(x,R(x)\right) \,.
\end{equation}
such that $g\xrightarrow{R\rightarrow 0} 0 $. In other words, under the above \emph{assumptions}, we can write
\begin{equation}\label{g_expand}
g\left(x,R(x)\right) = q R(x)+ \frac{h}{x}R(x)+... \,,
\end{equation}
where we have expanded for small $1/x$ and $R$, and kept only the leading terms. Since the RG functions $\gamma$ and $\beta$ are not independent of each other, a similar expression to Eq.~\eqref{gamma} must hold for the beta function:
\begin{equation}\label{beta}
\beta(x)= \beta_{pert}(x) + b\left(x,R(x)\right) \,.
\end{equation}
where, for the function $b\left(x,R(x)\right)$, the same logic of $g\left(x,R(x)\right)$ is valid. As stated at the beginning of this section, we are now considering nonperturbative corrections to the beta function and not only its asymptotic expansion.

The next step is to replace Eqs.~\eqref{gamma} and~\eqref{beta} in Eq.~\eqref{maineq}, expand for small $1/x$ and $R$, and rescale $R\rightarrow R/x$~\footnote{This manipulation is needed to bring the equation in the normal form because, unlike Ref.~\cite{Bersini:2019axn}, we are consistently considering also the nonperturbative correction to the beta function. This rescaling can also modify the kind of poles in the Borel plane and bring them in a convenient form~\cite{CostinBook}.}. By working the previous steps at the first order of perturbation theory, one finds
\begin{equation}\label{ODE_explicit}
R'(x) = - q\, s R(x) - s\,\left(g_1 + h - \beta_1 \right) \frac{R(x)}{x} + f(R^2, R \, x^{-2})  + \mathcal{O}(R^3|R \,x^{2})\,,
\end{equation}
which gives Eq.~\eqref{ODE} -- except the analytic terms -- with the constant $k_1$ written in terms of the expansion coefficients: $g_1$ that is the one-loop coefficient of $\gamma_1(x)$; $\beta_1$ that is the one-loop coefficient of $\beta(x)$; $h$ that comes from the expansion in Eq.~\eqref{g_expand}. One can rewrite Eq.~\eqref{ODE_explicit}  in the normal form (as a non-homogeneous equation) by performing a change of variables via formally small analytic terms, $\bar{R}(x)=R(x)-\mathcal{O}(1/x^{N+1})$, and then renaming $\bar{R}$ as $R$, with abuse of notation.

Notice that the expansion of the function $b\left(x,R(x)\right)$ in Eq.~\eqref{beta} enters only in the coefficient of $R^2$ (not explicitly written) or higher powers. Therefore,  albeit the nonanalytic corrections to the beta function have to be considered \textit{a priori}, their presence does not spoil either the position or the type of poles in the Borel transform of the function $R(x)$.

Notably, the parameter $q$ is fixed by matching with the Landau pole structure, i.e. from the nonanalytic dependence on $\alpha$ of the Green function as shown in Eq.~\eqref{rbis}, which determines the non-analyticity and thus the transseries structure of $R(x)$. As a result, one obtains
\begin{equation}\label{qs}
q=s = \text{sign}(\beta_1) = \pm 1 \,.
\end{equation}
A detailed resurgent analysis of Eq.~\eqref{ODE} (or \eqref{ODE_explicit}) shall be presented in Sec.~\ref{sec:resurgence}.

\paragraph{Remarks on the resurgent ODE.}
The  Eq.~\eqref{ODE} guarantees that renormalons can be resummed applying the isomorphism presented in Refs.~\cite{Costin1995,costin1998}. As a result, there are no longer infinite ambiguities due to the renormalons, and only one arbitrary constant needs to be determined by the initial condition for Eq.~\eqref{ODE}.  However,  renormalons do not have a semiclassical limit (unlike instantons). Therefore, it is impossible to determine the aforementioned arbitrary constant beforehand using QFT considerations.

It is important to stress that the leading term $- R(x)$ in Eq.~\eqref{ODE} determines the position of the poles in the Borel transform of the Green functions. In this work, we show that  $q$ in Eq.~\eqref{qs} is uniquely fixed from Eq.~\eqref{rbis}, with no need of matching with any skeleton diagram as in Ref.~\cite{Bersini:2019axn} -- see also the discussion in Subsec.~\ref{sub:complete}. Ultimately,  Eqs.~\eqref{qs} and~\eqref{rbis} are the consequence of the finite radius of convergence of the series~\eqref{PT_ren}.

The Eq.~\eqref{ODE} shows that the -- necessary -- nonperturbative completion of perturbation theory is resurgent and has the analytical properties of the renormalons, namely having singularities in the Borel plane as in the example of Eq.~\eqref{QEDrenexample}.
The parameter $k_1$ in Eq.~\eqref{ODE} determines the type of poles in the Borel transform of the Green functions. However, $k_1$ depends on the parameter $h$, which, in turn, comes from the expansion for small $R$ in Eq.~\eqref{g_expand}. Being this independent from the loop (perturbative) expansion, $h$ (and $k$) cannot be calculated in perturbation theory. The fact that the type of poles is inaccessible from perturbation theory can also be understood as follows: suppose one calculates some renormalon poles by evaluating skeleton diagrams~\footnote{It may help to stress that we call skeleton diagram the kind of diagram first proposed by 't Hooft~\cite{tHooft:1977xjm}, and not the primitive in the Hopf algebra language.} for the two-point Green function(e.g. Ref.~\cite{Neubert:1994vb}). Then, higher $n$-point Green functions can be found from the two-point Green function using the Schwinger–Dyson equations.
As first pointed out in Ref.~\cite{tHooft:1977xjm}, operations of differentiation and integration affect the type of poles but not their position. Consequently, the type of poles in the Borel transform of the $n$-point Green function differs from the original singularities in the two-point Green function. However, one can directly estimate the skeleton diagram contribution to the $n$-point Green functions, finding the same kind of singularities of the two-point Green function, evaluated similarly. The latter arguments suggest that the type of poles in the Borel transform of the Green functions cannot be reliably estimated from perturbation theory. Nevertheless, Eq.~\eqref{ODE} guarantees the applicability of the generalized Borel resummation of Ref.~\cite{Costin1995}. Hence, the analytic structure of  $R(x)$ can be extracted \textit{a posteriori} from experimental data.

\section{Resurgence and Alien Calculus for Non-linear ODE}\label{sec:resurgence}

The existence of the  (nonlinear in $R$) ODE discussed in the previous section suffices to treat the renormalons through resurgence theory~\cite{Costin1995,costin1998}, and this has been imported in QFT in Refs.~\cite{Maiezza:2019dht}.
This section aims to give a self-contained insight to the reader on why Eq.~\eqref{ODE}, and thus renormalons, are resurgent. In doing this, we reproduce one of the central results of Ref.~\cite{Costin1995} regarding the Alien Calculus. Connections between the general theory of resurgence with the Alien calculus in quantum theory and field theory can be found in Refs.~\cite{Aniceto:2013fka,Dorigoni:2014hea}. In what follows, we focus on the Alien calculus versus resurgence specifically within ODEs.

\subsection{Resurgent properties and Borel-Ecalle resummation}

We summarize the main elements needed to understand Alien Calculus. In this section, the function  $R(x)$ denotes a generic nonanalytic function with infinite many singularities in its Borel transform at $1,2,...,\infty$.

\emph{Definition 1:} lateral Borel summation $S_{\theta^{\pm}}$
\begin{equation}
S_{\theta^{\pm}} R(x) = \int_0^{e^{\pm i \theta(\infty\pm i\epsilon)}} \, dz \, e^{-xz} \, \mathcal{R}(z)
\end{equation}
For instance, in QED, the Borel transform $\mathcal{R}(z)$ features an infinite number of singularities along the real axis ($\theta=0$). The lateral Borel summations $S_{0^{\pm}}$ give different results in this case.

\emph{Definition 2:} The Stokes automorphism $G_{\theta}$ is given by
\begin{equation}
S_{\theta^{+}} \, R(x) = S_{\theta^{-}}  \circ  G_{\theta}\, R(x) = S_{\theta^{-}}  \circ  (1+\delta_{\theta})\, R(x),
\end{equation}
where  $\delta_{\theta}$ denotes the discontinuity along the singular line. The Stokes automorphism allows one to compute one lateral summation once the other is known.

\emph{Definition 3:} The alien derivative. One can write the Stokes automorphism $G_{\theta}$ as
\begin{equation}\label{stokesauto}
G_{\theta} = e^{\log G_{\theta}} := e^{\dot{\Delta}_{\theta}},\text{ where} \quad \dot{\Delta}_{\theta} = \log G_{\theta} = \log(1+\delta_{\theta}) =\sum_{n=0}^{\infty} \frac{(-1)^n}{n}\delta_{\theta}^n \,.
\end{equation}
Due to the morphism properties of $G_{\theta}$ in the multiplicative space, $\dot{\Delta}_{\theta}$ satisfies Leibniz's rule~\cite{Ecalle1993}
\begin{equation}
e^{\dot{\Delta}_{\theta}}(f\, g) = e^{\dot{\Delta}_{\theta}}(f) e^{\dot{\Delta}_{\theta}}(g) \,.
\end{equation}
The functions $f$ and $g$ denote two generic nonanalytic functions. Expanding the exponential $e^{\dot{\Delta}_{\theta}}$ leads
\begin{equation}
(1+ \dot{\Delta}_{\theta}+ \mathcal{O}(\dot{\Delta}_{\theta}^2) )f\, g= e^{\dot{\Delta}_{\theta}}(f) e^{\dot{\Delta}_{\theta}}(g) = f\, g+g \dot{\Delta}_{\theta} f + f \dot{\Delta}_{\theta} g + \mathcal{O}(\dot{\Delta}_{\theta}^2) \,.
\end{equation}
Comparing both sides, one finds
\begin{equation}
\dot{\Delta}_{\theta}(f\,g) = g \dot{\Delta}_{\theta} f + f \dot{\Delta}_{\theta} g\,.
\end{equation}
Analogously, linearity of the alien derivative, namely $\dot{\Delta}_{\theta}( \alpha f) = \alpha\dot{\Delta}_{\theta}(f)$ can be derived from the morphisms property $e^{\dot{\Delta}_{\theta}}(\alpha f) = \alpha e^{\dot{\Delta}_{\theta}}(f)$, where $\alpha$ is a complex constant. Another essential property of the alien derivative is that it commutes with the standard derivation~\cite{Ecalle1993,EcalleRes:book}:
\begin{equation}\label{aliencom}
\left[\dot{\Delta}_\theta , \frac{d}{dx}	\right]=0 \,.
\end{equation}
In what follows, we shall find the explicit expression for the alien derivative $\dot{\Delta}_{\theta}$ in the context of the equation shown in Eq.~\eqref{ODE}. For concreteness, we shall take $\theta=0$, which is the case relevant for QFT (specifically, for renormalons).\\

The solution of Eq.~\eqref{ODE}, for any analytic $f$, is a one-parameter transseries ($C$) of the form~\cite{CostinBook}
\begin{equation}\label{resol}
R(x)= \sum_{n=0}^{\infty}C^n e^{-n\, \, x}  \, R_n(x)  \,,
\end{equation}
$R_0(x)$ being the formal solution of Eq.~\eqref{ODE} at order $C^0$. In general, the function $R_0(x)$  has a divergent asymptotic expansion, and it is the source of higher-order renormalon contributions to the Green function in the QFT context. We want to show that once the function $R_0(x)$ is known, the other functions $R_1(x), R_2(x),...$ are obtained by repeatedly applying the alien derivative to $R_0(x)$.

To derive the resurgence properties of Eq.~\eqref{resol}, we apply $\partial_C R(x)$ and $\dot{\Delta}_0 R(x)$ to Eq.~\eqref{ODE}. Because of the commutation relation in Eq.~\eqref{aliencom}, one has
\begin{align}\label{brigde1}
\frac{d}{d x} \partial_{C} R(x)  & = - \partial_{C} R(x)+\frac{1}{x} k_1 \partial_{C}R(x)+\partial_{C} f\, , \\
\frac{d}{d x} \dot{\Delta}_0 R(x)& = -\dot{\Delta}_0R(x)+\frac{1}{x} k_1 \dot{\Delta}_0R(x)+\dot{\Delta}_0 f\,,
\end{align}
where we used the property that the Alien derivative of an analytic function is zero. Since the two equations are formally the same, it must be that the derivative with respect to $C$ and the alien derivative are proportional to each other, namely
\begin{equation}\label{bridgeeq}
\dot{\Delta}_0 R(x) = S(C) \partial_C R(x)\,.
\end{equation}
The Eq.~\eqref{bridgeeq} is known as Ecalle Bridge equation, and $S(C)$ is in general a function of $C$ that we expand as
\begin{equation}\label{bridgeeq2}
S(C) = \sum_{m=0}^{\infty} S_m C^m\,.
\end{equation}
We substitute Eq.~\eqref{resol} in the bridge equation~\eqref{bridgeeq} to derive the resurgence of the functions $R_n(x)$, namely in the summation below
\begin{align}
\dot{\Delta}_0R(x) & = \sum_{n=0}^{\infty}C^n e^{-n\, \, x}  \, \dot{\Delta}_0 R_n(x)\,, \label{a1}\\
S(C) \partial_C R(x) & = S(C) \sum_{n=1}^{\infty}n\, C^{n-1} e^{-n\, \, x}  \,  R_n(x)\nonumber \\
& = e^{-  x}\sum_{m=0}^{\infty} \sum_{n=0}^{\infty} S_m(n+1) C^{n+m} e^{-n  x} R_{n+1} \label{a2}  \,.
\end{align}
Both sides of Eqs.~\eqref{a1} and \eqref{a2} equal when
\begin{equation}
C^{n+m} e^{-n\, \, x} = C^n\,e^{-n\, \, x} \,
\end{equation}
which can only hold for $m=0$ and
\begin{equation} \label{bridgeeq3}
 \dot{\Delta}_0 R_n(x) =S_0 (n+1) e^{-  x} R_{n+1}(x)\,,
\end{equation}
where $S_0$ is called the holomorphic invariant, which is a pure imaginary number, since the alien derivative anti-commutes with complex conjugation and $R$ is real, thus
\begin{align}
& \left( \dot{\Delta}_0 R(x)\right)^* = \left(S_0\partial_C R(x)\right)^*\,,  \nonumber \\
& - \dot{\Delta}_0 R(x)  = S_0^*\partial_C R(x) = - S_0\partial_C R(x)  \label{complexalien} \,,
\end{align}
which implies $ S_0^* =- S_0$.

The Eq.~\eqref{bridgeeq3} gives the resurgence property of the transseries~\eqref{resol}. The  recursion in Eq.~\eqref{bridgeeq3} can be solved, and the functions $R_n(x)$ can be expressed as
\begin{equation}\label{solvedrecursion}
( \dot{\Delta}_0)^n R_0(x) = n!\, S_0^{\,n}\, e^{- n x} \, R_{n}(x)\,.
\end{equation}
Notice that Eq.~\eqref{solvedrecursion} is exact so far, but unfortunately, it is challenging to solve in practice since $\dot{\Delta}_0$ requires to evaluate infinitely many discontinuities. However, using the formal expansion for the alien derivative in Eq.~\eqref{stokesauto}, it is possible to rewrite $R_n(x)$ in Eq.~\eqref{solvedrecursion} in terms of $\delta_0 R_0(x), R_1(x), ..., R_{n-1}(x)$. This enables us to recast Eq.~\eqref{solvedrecursion} in the form
\begin{equation}\label{costin_mult}
R_n(x) = \frac{e^{n  x}}{S_0^{\,n}}\left(\delta_{0}R_0(x) -\sum_{j=1}^{n-1}S_0^{\,j}e^{-j x} R_j(x)	\right) \,\,\,, n\geq1 \,.
\end{equation}
The bottom line is that Eqs.~\eqref{resol} and~\eqref{costin_mult} are in agreement with the result obtained in Refs.~\cite{Costin1995,CostinBook} (see App.~\ref{app:resum} for further details). In realistic QFT applications, $R_0(x)$ can only be known approximately since all the terms in Eq.~\eqref{ODE_explicit} are not known. 

Finally, just taking the Cauchy principal value of the result shown in Eq.~\eqref{costin_mult} does not automatically preserve the homomorphism structure of the summation procedure for the transseries in Eq.~\eqref{resol}~\footnote{It turns out that Eq.~\eqref{costin_mult} can be made real using the fact that the holomorphic invariant is a purely imaginary number. However, in the more general case where reality and homomorphism are not automatically achieved, medianization preserves reality and the homomorphism structure of the summation procedure. }. To this end, the medianization (or balanced average) is defined  -- see appendix~\ref{app:generic} for more details. For our purposes, medianization is given by -- see Ref.~\cite{Dorigoni:2014hea},
\begin{equation}\label{med}
S^{med} := S_{0^{-}} \circ G_0^{1/2} =  S_{0^{+}} \circ G_0^{-1/2} \,,
\end{equation}
such that
\begin{equation}\label{reality}
R^{med}(x)=  S^{med} R(x) \,,
\end{equation}
where due to Eq.~\eqref{aliencom}, the function $R^{med}(x)$ is guaranteed to be a solution of Eq.~\ref{ODE}. The balanced average was defined in Refs.~\cite{Costin1995,CostinBook}, and because of the features of the ODEs, is a simplified form of the more general medianization defined in Refs.~\cite{Ecalle1993,EcalleRes:book}.

\subsection{On the Holomorphic Invariant $S_0$}

Because of the simple structure of the function, $S(C)$ in Eq.~\eqref{bridgeeq2}, the holomorphic invariant $S_0$  can be reabsorbed in $C$. To explicitly show this point, consider again the bridge equation~\eqref{bridgeeq}
\begin{equation}
\dot{\Delta}_0 R(x) = S(C) \partial_C R(x)= \left(  \sum_{m=-\infty}^{\infty} S_m C^m  \right) \partial_C R(x)\,.
\end{equation}
Since $C$ is arbitrary, one can always do the following change of variable
\begin{equation}\label{change_C}
C = S_0 C'\,,
\end{equation}
and thus
\begin{equation}
\dot{\Delta}_0 R(x) =  \left(  \sum_{m=0}^{\infty} S_m S_0^n C'^m  \right) \frac{1}{S_0}\partial_{C'} R(x)\,.
\end{equation}
Recall now that the only non-zero contribution to the above equation comes from $m=0$, thus it can be written as
\begin{equation}
\dot{\Delta}_0 R(x) =    S_0    \frac{1}{S_0}\partial_{C'} R(x)=     \partial_{C'} R(x)\,.
\end{equation}
Therefore, $S_0$ can be reabsorbed by a redefinition of the arbitrary constant $C$, and the first order Eq.~\eqref{ODE} has only one free constant, as it must be.

\section{Self-complete QED}\label{sec:QED}

From the discussion in Sec.~\ref{sec:RGE}, QFT has to be self-complete, or valid for any $L$, due to nonperturbative effects. The avoidance of the Landau pole in QED was previously investigated in Ref.~\cite{Klaczynski:2013fca}. The analysis of Ref.~\cite{Klaczynski:2013fca} is mainly based on the perturbative expression~\eqref{PT_ren}, with some \textit{ad hoc} ansatz on the recursion of the $\gamma_k$. The present paper shares with Ref.~\cite{Klaczynski:2013fca} the aspect that the nonperturbative or ``flat contributions" would allow the avoidance of the Landau pole.

We shall apply to QED the logic developed in Subsec.~\ref{sub:complete} together with the tools developed in Secs.~\ref{sec:ODE} and~\ref{sec:resurgence}. We partially follow Ref.~\cite{Maiezza:2020nbe} in which the authors apply Resurgence on top of renormalons to render self-complete the $\phi^4$ model. In the $\phi^4$ model, however, the relation between the beta and the gamma functions involves both the anomalous dimensions of two-point and four-point Green functions, making the avoidance of the Landau pole less transparent. In this work, we argue that avoiding the Landau pole is a necessity when $R(x)$ is included in Eq.~\eqref{complete}. Furthermore, here we provide a technical improvement concerning Ref.~\cite{Maiezza:2020nbe} since we consider a more generic pole structure for the Borel transform of the two-point Green function (not only simple poles). Branch points in the Borel transform of the Green function induce non-zero, higher nonperturbative sectors (i.e., the $R_n$ in Eq.~\eqref{resol}). We can also explicitly check that the series in $R_n$ in Eq.~\eqref{resol} can be safely truncated at some order $n=n_0$ for some sufficiently small critical value of the coupling constant.  \\

For QED, one has
\begin{equation}\label{QED_specific}
\beta(\alpha)= \alpha \gamma(\alpha)\,,
\end{equation}
which simplifies Eq.~\eqref{ODE_explicit} such that one has
\begin{equation}\label{QED_ODE}
R'(x)= - R(x) +\frac{R(x) (\beta_2-\beta_1 h)}{\beta_1 x} + \frac{R(x) (\beta_2 (\beta_1 h-\beta_2)+
\beta_1 R(x) x)}{\beta_1^2 x^2} +\text{analytic terms}  \,,
\end{equation}
where $\beta_1$ has been defined below Eq.~\eqref{conv}, $h$ is already known from Eq.~\eqref{g_expand}, and $\beta_2$ denotes the two-loop coefficient of the beta-function.

As argued in Sec.~\ref{sec:ODE}, perturbation theory has no access to some parameters -- e.g., $h$ in the coefficient of $\frac{R}{x}$ in Eq.~\eqref{QED_ODE}. Thus, perturbation theory cannot determine the kind of poles beforehand.  Notice, however, that Eq.\eqref{QED_ODE} formally determines both the position and the type of poles solely in terms of $\beta_1$, $\beta_2$ and $h$ which are scheme independent -- since $h$ not calculable in terms of loops and therefore scheme independent. Including higher loops and higher terms in Eq.~\eqref{g_expand} does not modify the terms shown in Eq.~\eqref{QED_ODE}.

In App.~\ref{app:generic}, we consider a generic pole structure for the Borel transform, which induces an infinite number of nonperturbative functions $R_n$ with $n\geq1$. The general treatment requires several technicalities discussed in Apps~\ref{app:resum} and ~\ref{app:generic}. For clarity, we illustrate in detail the renormalon resummation for simple poles. Let us consider the structure for the Borel transform of the Green function
\begin{equation}\label{sim_pole}
\mathcal{R}_0(z) = \sum_{n=1}^{\infty} (-1)^n\,\left( z- n  \right)^{-1}\,,
\end{equation}
in which the alternate sign is typical of any skeleton-diagrams estimate of renormalons structure~\cite{tHooft:1977xjm}.
Employing the isomorphism of Ref.~\cite{CostinBook} -- re-derived in the previous section -- one finds (in the original variable $\alpha$)
\begin{equation}\label{R_QED}
R^{med}(\alpha)=  R_0^{PV}(\alpha) - \frac{2\pi \,C}{1+e^{\frac{3\pi}{\alpha}}} \,,
\end{equation}
where $R_0^{PV}(\alpha)$ is the Cauchy principal value of the Laplace integral of Eq.~\eqref{sim_pole} (see Eq.~\eqref{reality} and appendix~\ref{app:generic})
\begin{equation}
R_0^{PV}(\alpha) = \text{P.V.} \left(\int_0^{\infty} dz e^{-\frac{3 \pi z}{\alpha}}\mathcal{R}_0(z)\right) \,.
\end{equation}
It has a power series expansion in $\alpha$, and we also include into $R_0^{PV}$ the low-order perturbative contributions.

By using Eqs.~\eqref{gamma},~\eqref{QED_specific} and~\eqref{R_QED}, the nonperturbative correction to the beta function are given by
\begin{equation}\label{final_beta}
\beta(\alpha)= \beta_{pert}(\alpha) - \alpha \frac{2\pi \,C}{1+e^{\frac{3\pi}{\alpha}}} \,,
\end{equation}
where  $\beta_{pert}$ denotes the analytic contributions coming from perturbation theory and the higher-order term $\alpha R_0(\alpha)$ in Eq.~\eqref{R_QED}.

\begin{figure}
 \centering
 \includegraphics[width=0.6\columnwidth]{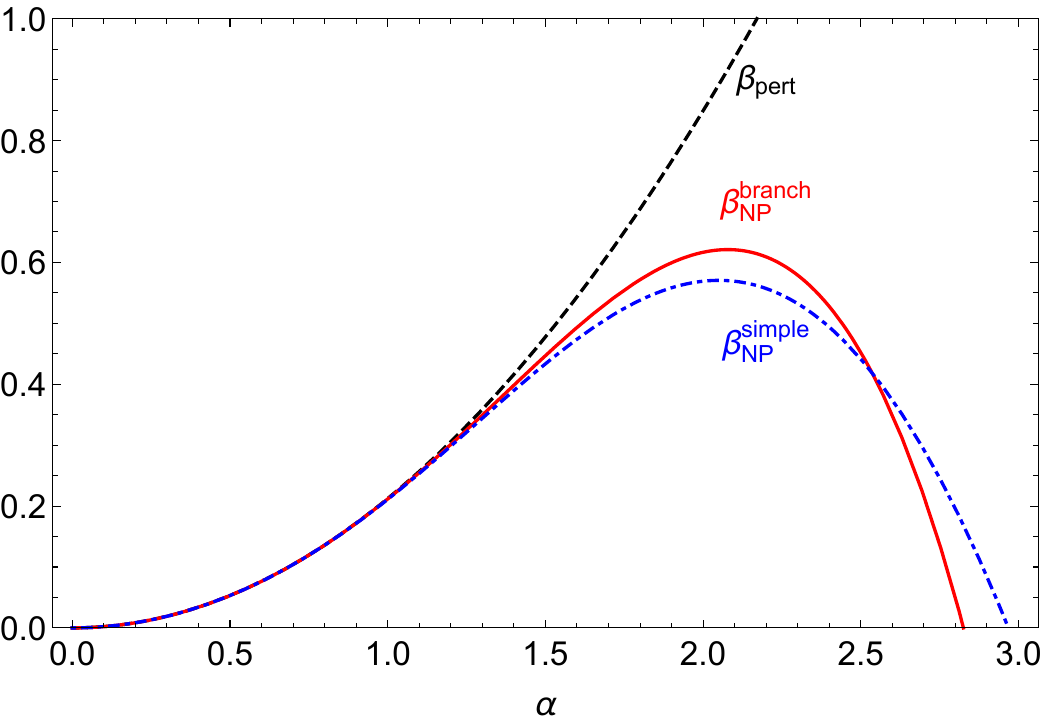}
 \caption{Comparison of the perturbative beta-function ($\beta_{pert}$, dashed, black line), the nonperturbative one ($\beta_{NP}^{simple}$, dot-dashed, blue line) for simple poles in Eq.~\eqref{final_beta}, and the nonperturbative one ($\beta_{NP}^{branch}$, plain, red line) for square-root branches as in App.\ref{app:generic}. The transseries constant $C$ is fixed to $C = 2.5$.}
 \label{beta_function}
\end{figure}

Notice that the nonperturbative contribution in Eq.~\eqref{final_beta} can be used to find a new nonperturbative zero of the beta function $\beta(\alpha_c)=0$. The critical value $\alpha_c$ is a function of \textit{a priori} arbitrary constant $C$. In Fig.~\ref{beta_function}, we show the beta function in the cases of simple poles and branch points structure of the renormalons contribution, compared to the perturbative, one-loop expression $\beta_{pert} \approx \beta_1 \alpha^2$.

From Fig.~\ref{beta_function}, we see that, as long as the coupling $\alpha$ is sufficiently small, a more general pole structure mildly impacts the numerical value of the critical point $\alpha_c$. The reason is that the higher nonperturbative sectors ($R_n$ with $n\geq 1$ ) are subleading for $\alpha\leq \alpha_c$, keeping the truncation of the higher sectors in App.~\ref{app:generic} under control. Notice that this example is not quantitatively reliable since we are considering only one-loop contributions; nonetheless, Fig.~\ref{beta_function} conceptually points out how nonperturbative contributions can impact the running of $\alpha$, making it asymptotically safe.\\

Some concluding remarks are in order:

\begin{enumerate}[label=(\alph*)]

\item  We are arguing on the presence of renormalons in the beta function -- the authors of the recent Ref.~\cite{Dunne:2021lie} discussed in detail this issue in the specific case of $\phi^4$ model. Here, we should recall that the absence of renormalons in the beta function is solely an artifact of the $\overline{MS}$ renormalization scheme. For example, within the on-shell renormalization scheme, which is considered more physical, the beta function has renormalon singularities~\cite{Broadhurst:1992si}.

\item Because of the nonexistence of a semiclassical limit of the renormalons, one cannot predict the value of $C$, which remains undetermined even from the requirement of asymptotic safety in Fig.~\ref{beta_function}. The constant $C$ in the two-point function will appear in any Green functions because of the Schwinger–Dyson equation. More important, it is non-Lagrangian, namely, it is not related to the Lagrangian (of QED, in our case). The latter suggests that, beyond perturbation theory and due to the renormalons, the standard Lagrangian field theory approach fails in the complete, fundamental description of a given QFT system. Therefore, renormalon singularities remain an illness of perturbative four-dimensional QFT~\footnote{It is worth mentioning that this insurmountable problem requires going beyond the standard setup of QFT. For example, appealing to somewhat exotic properties of spacetime, as in Refs.~\cite{Argyres:2012vv,Maiezza:2022xqv}, the issue can be avoided: in the former, the authors provide a semiclassical limit for IR renormalons in a QFT built on a compactified space; in the latter, the authors show that UV renormalons do not exist for a QFT augmented with the concept of dimensional reduction at high-energy scales.}. Moreover, there is also the uncomputable (from PT) parameter $h$ which enters in the determination of the kind of singularities, minimally affecting the result in Fig.\ref{beta_function}.
Regardless of this conceptual issue, the resurgence theory significantly improves the approach to renormalons: a few constants ($C$, $h$) are sufficient to describe them, instead of the infinite arbitrary constants introduced via operator-product expansion (e.g., Ref.~\cite{Beneke:1998eq}). Therefore, one can \textit{a posteriori} extract the value of $C$ in realistic models by measuring the running coupling at different energies or computing nonperturbative corrections proportional to $C$ to some physical observable and then matching with data. Alternatively, identifying the dynamic universality class of the critical point in the QFT system under consideration may be possible so that the values of the critical exponents are known. Consequently, the constant $C$ might be determined by universality considerations (this applies for models where the IR renormalons are the ones to lie on the semi-positive axis, so not in the case of QED). For example, in this regard, see Ref.~\cite{Son:2004iv}, in which the authors suggest the universality class for QCD to be that of a model H, i.e., the liquid-gas phase transition.

\item  From the argument presented in this paper, we have seen that renormalons already appear within leading order calculations, making one think that they are the leading singularities in the Borel plane, as initially argued in Ref.~\cite{Parisi:1978iq}. There is a region where the coupling is sufficiently small, and one can neglect instanton corrections. However, as the coupling increases, the instantons become increasingly important and cannot be ignored. The way to treat instantons together with renormalons is an open problem. One challenging difficulty lies in the fact that there is an overlap of two Stokes lines on the semi-positive axis, one due to the renormalons and the other due to the instantons. This phenomenon is known as ``resonance" (e.g., see Ref.~\cite{CostinBook}). Nevertheless, as long as the critical coupling $\alpha_c$ is smaller than the position of the first instanton singularity in the Borel transform of the Green functions, there is the possibility that instantons remain practically irrelevant, such that one may circumvent the resonance problem.

\item The QED is intrinsically a part of the Standard Model. Given this, one would anticipate that the recognized weak force corrections would play a role before the nonperturbative effects, known as renormalon effects, become important. This anticipation stems from the relatively low value of $\alpha_{QED}\approx1/137$ at energies below the electro-weak scale. As a result, renormalon contributions are drastically reduced, approximated by $\approx e^{-1/(\beta_1 \alpha_{QED})}$. Similar reasoning can be applied to Fig.~\ref{beta_function}, which illustrates a UV fixed point, signifying asymptotic safety for QED. However, in a practical scenario, this would occur well beyond the electro-weak and even the Planck scale. The presence of intermediate scales does alter the dynamics. Still, our primary theoretical argument aligns with the U(1) gauge model prototype.
Broadly speaking, these ideas can be expanded to more comprehensive models of Quantum Field Theory (QFT), including the Standard Model. One challenge here might be the multitude of couplings. A tangible example where our theory could be applied is in Quantum Chromodynamics (QCD). Here, the emphasis is on the infrared region, but the underlying math remains consistent.

\end{enumerate}

\section{Discussion and Summary}\label{sec:summary}

\paragraph{Wilsonian renormalization and Resurgence}
It may be interesting to compare the resurgent approach to renormalization versus the Wilsonian one.
Wilson's theory assumes as a starting point that any QFT has an intrinsic energy scale $\Lambda_{cutoff}$ at which the description ceases to work~\cite{Wilson:1973jj}. Therefore, the cutoff is endowed with a physical significance in this approach. For example, the cutoff requirement would imply that spacetime becomes discrete at lengths $\propto1/\Lambda_{cutoff}$. To make the continuum limit of an interacting QFT, one has to send the cutoff to infinity, meaning that one has to find non-Gaussian fixed points of the beta function~\footnote{The recent Ref.~\cite{Romatschke:2023sce} discusses the continuum limit for $\phi^4$ model.}. This has been known since the seminal work of Wilson~\cite{Wilson:1973jj} (see also Ref.~\cite{Peskin:1995ev}, chapter 12), and we have shown that Resurgence provides a framework to find a new kind of fixed points.

However, our approach is orthogonal to the Wilsonian one: we start from the principle that RGE flow is well-defined at any energy for a model proclaimed as fundamental -- such as QED, and then we find the analytic structure of the nonperturbative completion. The reader may worry about our principle in the light of known supersymmetric models where the beta functions are calculable ``exactly" at one loop using chiral anomalies and features a Landau pole. However, the ``exact" one-loop computation disagrees with the standard perturbation theory evaluations of the beta function receiving higher-order contributions ~\cite{Avdeev:1981ew}. The author of Ref.~\cite{Shifman:1986zi} proposed to solve this issue by distinguishing between the Wilsonian and the physical, one-particle-irreducible coupling. The latter can then have nonperturbative corrections consistent with what we have discussed in our analysis.

\paragraph{Summary}
 With the nonperturbative nature of the renormalization group equation in mind, we have first given a generic argument on why the perturbative expansions in the four-dimensional QFT must be divergent in some limit, independently of the instantons. We have shown that the two-point function logarithm expansion in Eq.~\eqref{PT_ren} (or scale expansion) has a finite radius of convergence, thus manifestly failing to describe the correlator for large logarithms. The latter is related to the Landau pole (at the ultraviolet or infrared) concept that we rephrased in the context of resurgence theory.

Because of the above principle, the next step has been to introduce a required nonperturbative function to complete the logarithm expansion -- see Eq.~\eqref{complete} to have a renormalization group flow well-defined at any energy. Improving several results from the literature, we have then shown that this nonperturbative function has to be identified with the Borel-Ecalle resummation of the renormalons. This proves that the nonperturbative function obeys a specific, nonlinear ordinary differential equation shown in Eq.~\eqref{ODE_explicit}. Thus, starting with a few hypotheses, we have proved the appearance of singularities located at $z= 2 n/\beta_1$, so identifiable with the usual renormalons, with no direct reference to skeleton diagrams (the only diagrammatic input is the calculation of $\beta_1$). On the contrary, we have argued that the skeleton diagrams cannot further shed light on the differential equation describing the renormalons.

Common lore, based on perturbation theory, is that there is a correspondence between the Landau pole and renormalons. This is an artifact and, in the proposed resurgent framework, one indeed sees the opposite: one defines a nonperturbative function on top of the perturbative expansion, such that QFT remains defined at all energy scales, then one shows that it is in correspondence with the renormalons. Taking as a prototype the QED, we have explicitly shown to the leading order in both the coupling $\alpha$ and the nonperturbative correction, that nonperturbative corrections calculated using Resurgence can make an asymptotically safe theory. The latter agrees with the starting point of a well-defined renormalization group equation at all energies.
The perturbative Landau pole is not a cutoff but merely a bookkeeping scale at which the perturbative treatment ceases to work. Consequently, QED may not need an ultraviolet completion in terms of new degrees of freedom. In the light of the proposed resurgence theory in QFT, the effective-field-theory treatment of QFT is a possibility but not a necessity.

In all this, resurgence theory is a fundamental tool, enabling us to rephrase the description of the renormalons in a way inaccessible to perturbation theory.  As pointed out in Ref.~\cite{tHooft:2021itz}, within perturbation theory, ``we still do not quite understand what
the equations are at the most fundamental level''.  In this respect, Resurgence provides a framework that at least allows one to extract nonperturbative information from first principles from the experiments. Remarkably, along this line, we have shed light on the connections between the theory of the generalized Borel resummation based on the no-linear, ordinary differential equations and the Alien Calculus.

Finally, we comment on possible applications and consequences in the infrared. What happens if one applies our approach to a Yang-Mills theory? One has a similar situation to the QED case, namely eliminating the (infrared) Landau pole. However, we speculate that the Resurgence of the renormalons at the infrared might also provide insights into the confinement. The issue deserves a detailed investigation that we leave for future work.


\section*{Acknowledgement}

AM thanks the Department of Physics at Universit\'a dell’Aquila for stimulating comments. JCV thanks Jagu Jagannathan and colleagues at Amherst College Department of Physics and Astronomy for illuminating discussions.


\appendix


\section{Technical details on ODEs}\label{app:ODEsdetail}

This appendix aims to provide technical details on ODEs omitted in the main text.
\vspace{1em}

\paragraph{Proving the Eq.~\eqref{ratio}.}

Start from the system of ODEs~\eqref{ODE_system} and solve the $\gamma_k$, for any $k$, as a function of only $\gamma$, $\beta$ and their derivatives (with respect of $\alpha$):
\begin{align}
& \gamma _1(\alpha ) = \gamma (\alpha ) \nonumber \\
& \gamma _2(\alpha ) = \frac{1}{4} \left(\beta (\alpha ) \gamma '(\alpha )-2 \gamma (\alpha)^2\right) \nonumber \\
& \gamma _3(\alpha ) = \frac{1}{24} \left(\beta (\alpha ) \left(\beta '(\alpha ) \gamma '(\alpha)+\beta (\alpha ) \gamma ''(\alpha )\right)-6 \beta (\alpha ) \gamma (\alpha ) \gamma'(\alpha )+4 \gamma (\alpha )^3\right) \\
& ... \nonumber
\end{align}
Next, we replace in the above expressions the power series approximation for $\gamma$, $\beta$
\begin{align}
& \gamma \sim \sum_{n=1}^{\infty} g_n \alpha^n \\
& \beta \sim \sum_{n=1}^{\infty} \beta_n \alpha^{n+1}\,, \nonumber
\end{align}
and keep the leading expression in $\alpha$. This yields
\begin{align}
& \gamma _1(\alpha ) =  \alpha  g_1                     \nonumber\\
& \gamma _2(\alpha ) =  \frac{1}{4} \alpha ^2 g_1 \left(\beta _1-2 g_1\right)                                  \nonumber \\
& \gamma _3(\alpha ) =  \frac{1}{12} \alpha ^3 g_1 \left(\beta _1-2 g_1\right) \left(\beta _1-g_1\right) \,. \\
& ... \nonumber
\end{align}
Finally, considering the ratios $\gamma_1/\gamma_{2}$, $\gamma_2/\gamma_{3}$, $...$, and extrapolating $\gamma_k/\gamma_{k+1}$ for any $k$ gives Eq.~\eqref{ratio}:
$$
\frac{\gamma_k(\alpha)}{\gamma_{k+1}(\alpha)} \simeq \frac{1}{\alpha \left( \frac{k \beta_1}{2(k+1)}- \frac{g_1}{k+1} \right)}\,.
$$
Eq.~\eqref{r} follows in the limit $k\rightarrow \infty$.

\paragraph{Non-linear ODE in the convolutive model.}

Now, we study the features of Eq.~\eqref{ODE} in the convolutive model.

\begin{enumerate}

   \item  First apply the Borel transform  to  Eq.~\eqref{ODE}, which gives
\begin{equation}
-z B(R(x)) = F_0(z) - B[R(x)] +k_1 \int_0^z B[R](z-z_1) dz_1 + \mathcal{O}(B[R(x)]^2)
\end{equation}
the $F_0(z)$ is the Borel transform of the higher-order analytic terms, and its precise form is not important for the following analysis. The above equation is then solved recursively for formally small $k_1$ and non-linear terms in $R$. So, starting with $k_1$ and the non-linear terms in $R$ zero, one has
\begin{equation}\label{A1}
B[R(x)] = \frac{F_0}{1-z}\,.
\end{equation}
This already shows that the Borel transform $B[R(x)]$ has a singularity at 1 in units of $2/\beta_1$ (since we use the change of variables of Eq.~\eqref{change_var} to normalize the position of the singularities in the Borel transform).

\item The next step consists in understanding the effect of the non-linear term $R(x)^2$ or higher in Eq.~\eqref{ODE}. For example, the Borel transform of a non-linear, quadratic term will result in a term in the Borel transformed Eq.~\eqref{ODE} of the type
\begin{equation}\label{A2}
\int_0^{z}B[R] (z-z_1) B[R] (z_1) dz_1\, .
\end{equation}
Then, if one plugs Eq.~\eqref{A1} into Eq.~\eqref{A2}, it is easy to convince oneself that one gets a branch point at $z=1$ and a new simple pole at $z=2$. Then, by doing this self-convolutions recursively, one finds an infinite number of poles
\begin{equation}
B[R(x)] = \sum_i \frac{F_0}{(z_i-z)}+ ...\,,
\end{equation}
at $z_i=1,2,3,...$ in the Borel transform $B[R(x)]$ and the $...$ is due to additional logarithmic branch points at $z_i$ which would give formally smaller contributions to $R$, when applying the Laplace transform.

\item
The final step includes recursively the effects of the term proportional to $k_1$. A straightforward calculation gives that the net effect of the $k_1$ term is to change the type of pole in $B[R(x)]$ from a simple pole to a general singularity of the form
\begin{equation}
B[R(x)] = \sum_i\frac{F_0}{(z_i-z)^{1+k_1}}\,.
\end{equation}

\end{enumerate}

\section{On the resummation isomorphism from the bridge equation}\label{app:resum}

The Eq.~\eqref{costin_mult} follows from Eq.~\eqref{solvedrecursion} by linearizing in $\delta_\theta$. Let us explicitly show this statement.

Replacing Eq.~\eqref{stokesauto} (with $\theta=0$)
\begin{equation}
\dot{\Delta}_0 = \sum_{n=0}^{\infty} \frac{(-1)^{n+1}}{n}\delta_{0}^n \,,
\end{equation}
into Eq.~\eqref{solvedrecursion}, one has
\begin{equation}
\left( \sum_{n=0}^{\infty} \frac{(-1)^{n+1}}{n}\delta_{0}^n \right)^k R_0 =  k! S_0^k e^{-k  x} R_k \,.
\end{equation}
As an example, let us write this equation up to $k=4$:
\begin{align}
& -\frac{1}{4}\delta_{0}^4 R_0 +\frac{1}{3} \delta_{0}^3R_0 -\frac{1}{2}\delta_{0}^2 R_0 +\delta _{0 }R_0   = R_1 S_0 e^{- x} \nonumber \\
& \frac{11}{12} \delta_{0 }^4R_0 -\delta_{0 }^3R_0 + \delta_{0}^2R_0  = 2 R_2 S_0^2 e^{-2   x} \nonumber \\
& \delta_{0}^3R_0 -\frac{3}{2}\delta_{0}^4 R_0   = 6 R_3 S_0^3 e^{-3   x}  \nonumber \\
& \delta_{0 }^4 R_0  = 24 R_4 S_0^4 e^{-4   x} \label{system} \,,
\end{align}
which is a linear system of four equations for four unknowns, $\delta_{0}^2 R_0\,, \delta_{0}^3 R_0\,, \delta_{0}^4 R_0, R_4(x)$. As a result, we obtain for $k=4$ the following:
\begin{equation}\label{explicit_costin}
R_4 = \frac{e^{4   x} \left(\delta_0  R_0 -R_1 S_0 e^{-  x} -R_2 S_0^2 e^{-2   x} - R_3 S_0^3 e^{-3  x}  \right)}{S_0^4} \,,
\end{equation}
which is the same  result of  Eq.~\eqref{costin_mult} for $k=4$. This holds for any $k$.
In particular, if we truncate to the leading correction $\delta_0 R_0$ for $R_1(x)$, it follows that
\begin{align}
& R_1(x) = \frac{e^{x} \left(\delta_0  R_0     \right)}{S_0} \,,  \\
&  R_n(x)=0 \,, \forall n \geq 2 \,.
\end{align}
Finally, applying the Borel transfrom to Eq.~\eqref{explicit_costin}, or the general Eq.~\eqref{costin_mult} gives  the equation (5.116) of Ref.~\cite{CostinBook} -- together with equation (5.117). In particular, the exponentials $\exp(k  x)$ give the operators $\tau_k$; our functions $R_n$ are denoted there as $y_n$, in the multiplicative model, while they become $Y_n$ in the convolutive one.

\section{Resummation for a generic kind of poles}\label{app:generic}

Consider the renormalon singularities in the convolutive model in the form
\begin{equation}\label{gen_pole}
\mathcal{R}_0(z)=\sum_{n=1}^{\infty} (-1)^n\,\left( z- n  \right)^{a-1}\,
\end{equation}
generalizing Eq.~\eqref{sim_pole} and being $0<a<1$. The latter can always be chosen by properly rescaling the nonlinear ODE $R\rightarrow R/x^s$, being $s$ some powers (see Ref.~\cite{Costin1995}). Applying any times the alien derivative, or the discontinuity operation $\delta_0$ in the system~\eqref{system}, to Eq.~\eqref{gen_pole}, one gets functions whose, in turn, have nontrivial discontinuities. This contrasts with the simple pole (i.e., $a=1$), where the operation $\delta_0$ gives an analytic function. Thus, in the generic case, one has infinite nonperturbative sectors proportional to powers of $C$. This is the meaning of Eq.~\eqref{costin_mult}, or the system~\eqref{system} for any $k$.

To be explicit, let us fix $a=1/2$ in Eq.~\eqref{gen_pole} and solve the Eqs.~\eqref{system} truncated at $k=4$ (i.e., four nonperturbative sectors), taking into account that
the p$^{th}$ discontinuity $\delta_0^p$, in the multiplicative model, reads
\begin{equation}\label{gen_dis_mult}
\delta_0^p\,R_0^{\pm}(x) = 2 (\pm i)^p \sum_{n=1}^{\infty} (-1)^n \sqrt{\frac{\pi}{x}} e^{- (n+p-1) x} \,
\end{equation}
where $R_0^{\pm}(x)$ denotes the upper and lower analytic continuation of $R(x)$ with respect of the Stokes line. The next step is to perform Ecalle's medianization, which reads as
\begin{equation}\label{realityapp}
R_i^{med}(x)=  S^{med} R_i(x) \,,
\end{equation}
with $i=1,2,3,4$ as in Eq.\eqref{system}. The imaginary factor $i$ in Eq.~\eqref{gen_dis_mult} is canceled by the holomorphic invariant $S_0$ in Eq.~\eqref{complexalien}.  In practice, recalling Eq.~\eqref{change_C}, the latter is equivalent to reabsorbing the $i$ in the constant $C$. Thus, the final resummed result is real, as it must be.

Finally, after summing over $n$, we obtain (in the variable $\alpha$)
\begin{align}
 R^{med}(\alpha) & = R^{med}_0(\alpha) + \sum_{k=1}^4C^k R^{med}_k(\alpha) e^{-\frac{3 \pi k}{\alpha}} + ... \nonumber \\ &
= \text{P.V.} \left(\int_0^{\infty} dz e^{-\frac{3 \pi z}{\alpha}}\mathcal{R}_0(z)\right)  + \frac{\sqrt{\alpha }}{e^{\frac{3 \pi }{\alpha }}+1} \left[ \left(2 \pi  e^{-\frac{3 \pi }{\alpha }}-\frac{11}{3} \pi ^{3/2} e^{-\frac{6 \pi }{\alpha }}+7 \pi ^2 e^{-\frac{9 \pi }{\alpha }}-2 \sqrt{\pi }\right) C \right. \nonumber \\
& \left(-2 \pi  e^{-\frac{3 \pi }{\alpha }}+4 \pi ^{3/2} e^{-\frac{6 \pi }{\alpha }}-\frac{25}{3}
   \pi ^2 e^{-\frac{9 \pi }{\alpha }}\right) C^2 + \left(4 \pi ^2 e^{-\frac{9 \pi }{\alpha }}-\frac{4}{3} \pi ^{3/2} e^{-\frac{6 \pi }{\alpha
   }}\right) C^3 \nonumber \\
& \left. - \frac{2}{3} \pi ^2 e^{-\frac{9 \pi }{\alpha }} C^4 +\mathcal{O}(C^5) \right] \label{gen_R}
\end{align}
where P.V denotes the Cauchy principal value, due to Eq.~\eqref{realityapp}, and we used the proper value of $\beta_1$ for QED.
The expression~\eqref{gen_R}, for square-root branch points, replaces the one for simple poles in Eq.~\eqref{R_QED}, and the rest conceptually remains the same.

\bibliographystyle{jhep}
\bibliography{biblio}

\end{document}